\documentclass[12pt,preprint]{aastex}

\begin{document}

\title{Keck Interferometer Observations of FU Orionis Objects}

\author{
R.~Millan-Gabet\altaffilmark{1},
J.~D.~Monnier\altaffilmark{2},
R.~L.~Akeson\altaffilmark{1},
L.~Hartmann\altaffilmark{3}, 
J.-P.~Berger\altaffilmark{4},
A.~Tannirkulam\altaffilmark{2},
S.~Melnikov\altaffilmark{5},
R.~Billmeier\altaffilmark{2},
N.~Calvet\altaffilmark{3}, 
P.~D'Alessio\altaffilmark{6},
L.~A.~Hillenbrand\altaffilmark{7},
M.~Kuchner\altaffilmark{8}, 
W.~A.~Traub\altaffilmark{3},
P.~G.~Tuthill\altaffilmark{9}, 
C.~Beichman\altaffilmark{1},
A.~Boden\altaffilmark{1},
A.~Booth\altaffilmark{10}, 
M.~Colavita\altaffilmark{10},  
M.~Creech-Eakman\altaffilmark{10}, 
J.~Gathright\altaffilmark{11},
M.~Hrynevych\altaffilmark{11}, 
C.~Koresko\altaffilmark{1},
D.~Le~Mignant\altaffilmark{11},
R.~Ligon\altaffilmark{10}, 
B.~Mennesson\altaffilmark{10}, 
C.~Neyman\altaffilmark{11}, 
A.~Sargent\altaffilmark{7},
M.~Shao\altaffilmark{10},
M.~Swain\altaffilmark{4},
R.~Thompson\altaffilmark{1},
S.~Unwin\altaffilmark{10},
G.~van~Belle\altaffilmark{1}
G.~Vasisht\altaffilmark{10}, 
\& P.~Wizinowich\altaffilmark{11}
}

\email{rafael@ipac.caltech.edu}

\altaffiltext{1}{Michelson Science Center, California Institute of Technology, MS 100-22, Pasadena, CA 91125}
\altaffiltext{2}{University of Michigan Astronomy Department, Ann Arbor, MI 48109-1090, USA.}
\altaffiltext{3}{Harvard-Smithsonian Center for Astrophysics, Cambridge, MA 02138, USA}
\altaffiltext{4}{Laboratoire d'Astrophysique de Grenoble, 414 Rue de la Piscine 38400 Saint Martin d'Heres, France}
\altaffiltext{5}{Ulugh Beg Astronomical Institute, Tashkent 700052, Uzbekistan}
\altaffiltext{6}{Universidad Nacional Aut\'{o}noma de M\'{e}xico}
\altaffiltext{7}{Astronomy Department, California Institute of Technology, Pasadena, CA}
\altaffiltext{8}{Princeton University, Princeton, NJ}
\altaffiltext{9}{University of Sydney, Physics Department}
\altaffiltext{10}{Jet Propulsion Laboratory, California Institute of Technology, Pasadena, CA 91109}
\altaffiltext{11}{W. M. Keck Observatory, California Association for Research in Astronomy, Kamuela, HI 96743}

\begin{abstract}

\end{abstract}

We present new K-band long baseline interferometer observations of
three young stellar objects of the FU~Orionis class, V1057~Cyg,
V1515~Cyg and Z~CMa-SE, obtained at the Keck Interferometer during its
commissioning science period. The interferometer clearly resolves the
source of near-infrared emission in all three objects.  Using simple
geometrical models we derive size scales (0.5$-$4.5~AU) for this
emission.  All three objects appear significantly more resolved than
expected from simple models of accretion disks tuned to fit the
broadband optical and infrared spectro-photometry. We explore
variations in the key parameters that are able to lower the predicted
visibility amplitudes to the measured levels, and conclude that
accretion disks alone do not reproduce the spectral energy
distributions and K-band visibilities simultaneously. We conclude that
either disk models are inadequate to describe the near-infrared
emission, or additional source components are needed.  We hypothesize
that large scale emission (10s of AU) in the interferometer field of
view is responsible for the surprisingly low visibilities.  This
emission may arise in scattering by large envelopes believed to
surround these objects.


\section{Introduction}

FU~Orionis objects are a small but remarkable class of stars that
display several-magnitude outbursts in visible light, followed by
decade-long fading phases.  They also exhibit fluxes in excess of
photospheric levels at infrared wavelengths, broad and doubled
spectral absorption lines, and wavelenght-dependent rotational
velocities. Two main theories compete to explain their
nature. \citet{herbig2003} (and references therein) favor a model in
which the flare occurs in an unstable star rotating near breakup, and
where the spectral properties are explained by a rapidly rotating
G$-$supergiant photosphere overlaid with a rising cooler shell.  In
contrast, a more widely accepted view first proposed by
\citet{hartmann1985}, is that FU~Orionis stars are pre-main sequence
objects and that all the observables have their origin in Keplerian
circumstellar disks and pre-natal infalling envelopes that surround
them \citep[see also][for recent arguments in relation to the disk
vs. rapidly rotating star debate]{hartmann2004}.  In this picture, the
outbursts have their origin in sudden increases in the accretion rate
through the disk, and all young stars may experience (probably
recurrent) FU~Orionis phases during their evolution \citep[see also
the review by][]{hartmann1996}.

Although the existence of circumstellar disks around young stars is
well established observationally, very little is directly known about
the structure and properties of the inner-most disk regions ($\lesssim
1$~AU). For solar-mass and intermediate mass young stars, disk models
have been reasonably successful at reproducing the infrared flux
excesses characteristic of these systems, as well as the spectroscopic
signatures of the accretion process itself. However, these efforts are
fundamentally limited by the fact that the observations are spatially
unresolved, which introduces important degeneracies with respect to
the spatial distribution of the emitting material.

The degeneracies with respect to the spatial distribution of material
inherent in spatially unresolved observations can be lifted by
observations using a long baseline interferometer, and important
progress has already been made using this technique.  For Herbig~Ae/Be
and T~Tauri objects (young stars of intermediate and solar-type mass
respectively), all observations to date have revealed surprisingly
large spatial scales for the near-infrared (NIR) emission
\citep{rmg1999,tuthill1999,rla2000,rmg2001,rla2002,eisner2004,rla2005,eisner2005,rla2005b,jdm2005};
prompting in part a significant revision of disk models, whereby
previously ignored detailed physics of the disk inner edge are now
incorporated with some success \citep{natta2001, dullemond2001,
jdm2002, muzerolle2003, vinkovic2003}. For a review of
observational breakthroughs in this area recently made possible by
interferometric techniques at infrared wavelengths, the reader is also
referred to \citet{rmg2006}.

For FU~Orionis objects, far fewer direct observational constraints
exist at those scales. On the other hand, following the initial
outburst the disk emission in FU~Orionis objects outshines that of
their central stars by many orders of magnitude, making them in
principle ideal laboratories for testing theories describing the
structure and physical properties of accretion disks. Indeed, the
interpretation is freed from several complicating factors that affect
the more evolved Herbig~Ae/Be and T~Tauri systems: instead of having
multiple source components (star plus disk) and disk heating
mechanisms (stellar irradiation plus active accretion), only the disk
component and active accretion matters for interpreting observations
of FU~Orionis objects.  Given that these same theories are being
applied to the disks inferred to exist around all young stars, and
that the physical conditions in the inner parts of those disks set the
initial conditions for planet formation, it is important to validate
the disk paradigm using FU~Orionis objects as ground truth tests.

The very first young stellar object to be resolved by long baseline
optical interferometry was in fact FU~Orionis itself, and in their
original analysis and follow up work \citet{malbet1998,malbet2005}
concluded that the visibility amplitudes measured agreed well with
those predicted by thermal emission in an accretion disk with the
canonical temperature profile $T \sim r^{-3/4}$ with parameters tuned
to fit the spectral energy distribution (SED).

In this paper we present new interferometer data on three other well
known FU~Orionis objects, and find instead that the NIR emission is
over-resolved relative to predictions of models that consist only of a
standard disk emitting thermally.  In $\S$~\ref{obs_sec} we present
the observations, instrumental arrangement and data analysis; in
$\S$~\ref{sizes_sec} we derive spatial scales for the NIR emission
using simple geometrical models of general applicability; in
$\S$~\ref{models1_sec} we construct accretion disk models with which
we attempt to simultaneously fit the broadband SEDs and visibility
amplitude data; in $\S$~\ref{models2_sec} we consider additional
source components that may be required in order to explain the
measured visibilities; and in $\S$~\ref{conclusions} we discuss our
results in the broader context and suggest future lines of
investigation.

\section{Observations and Data Analysis} \label{obs_sec}

We have observed three well studied FU~Orionis objects: V1057~Cyg,
V1515~Cyg and the South-East component of the binary system
Z~CMa. Basic properties of these objects are given in
Table~\ref{targets_tab}. We note that the photometry data given in
this table is meant to inform about the instrumental conditions only;
photometry used for modelling purposes was constructed as described in
$\S$~\ref{sed_sec}.

These observations were made at the Keck Interferometer (KI) during
its visibility science commissioning period (2002-2004). The KI
\citep{colavita2004} is formed by the two 10-m aperture Keck
telescopes separated by 85~m along a direction $\sim$38~degrees East
of North, corresponding to a minimum fringe spacing of
5.3~milli-arcseconds (mas) at 2.2$\mu$m.  In order to coherently combine the
NIR light from such large apertures, each telescope utilizes a natural
guide star adaptive optics system \citep{wiz2003}.  Optical delay
lines correct for sidereal motion and the telescope beams are combined
at a beamsplitter before the light is focused onto single-mode
(fluoride) fibers which impose a $\sim$50~mas (FWHM)
field-of-view (FOV) for all data reported herein.  While both H and
K-band observations are now possible, only broad K-band (2.18$\mu$m,
$\Delta\lambda= 0.3 \mu$m) data are reported here. For the four nights
over which the data presented here were taken, observing conditions
could be characterized as average, with fringe phase noise
(``jitter'', a measure of atmospheric seeing $+$ instrument noise) in
the range $0.5-1.25$~rad, and constant for calibrators and
targets. The system visibility, observed on calibrator point sources,
was always normal ($V^2_{system} = 0.6-0.7$).

We note that, although the angular separation of the components in the
Z~CMa system is only $\sim 0.1 \arcsec$ \citep{koresko1991}, the
adaptive optics sub-systems at KI clearly resolve the system, such
that either binary component can be selected for interferometer
measurements. Therefore, for the data presented here, the instrument
has isolated the contribution from the South-East component, believed
to be a FU~Orionis object \citep{hartmann1989,koresko1991}.  The
North-West component is believed to be a high-mass young star
surrounded by a dusty envelope \citep[see e.g.][]{garcia1999} and was
also resolved by KI; these results are discussed in \citet{jdm2005}.
With this method, and given the relatively small angular distance
between the two components, it is important to be sure that (1) the
fringe detector was pointed at the intended component, and (2)
received no extra flux from the other one. We have performed several
checks to verify this. First, the mean K-band fluxes measured from
each component are in a ratio (the Herbig component being $\sim \times
3$ brighter than the FU~Ori component) that approximately agrees with
(variable) values found in the literature (see references
above). Second, we have verified that the fringes for each component
were found at different optical delays, which differ by amounts
consistent with the $\sim 0.1 \arcsec$ astrometric offset. Finally,
the single-mode fibers which feed the fringe camera impose a FOV
($\sim 50$~mas as mentioned above) with an approximately Gaussian
tapering.  Therefore, even if assuming a $\times 3$ higher flux on the
(off-axis) Herbig component, its flux contribution to the on-axis beam
is less than 1\%, contributing negligibly to the measured
visibilities. Moreover, the single-beam fluxes fluctuate by 30\% or
less, very consistently for all objects (the ZCMa components and the
three calibrators used), indicating that uncompensated tip-tilt angle
fluctuations are much less than one FWHM, and again implying a small
fraction of flux contamination by ZCMa-NW to the ZCMa-SE measurement.

The square of the fringe visibility (V$^2$) was measured using the
ABCD-method \citep[using a dither mirror; see also][]{shao1977} and we
followed well-tested strategies described for the Palomar Testbed
Interferometer \citep{colavita1999}, except that corrections for
unbalanced telescope fluxes were improved, and jitter corrections were
not applied. These forms of jitter corrections are relevant when
optical path fluctuations are dominated by atmospheric residuals
outside the fringe tracker bandwidth. At KI, these fluctuations are
dominated by air motions in the Coude path and do not obey the same
statistics, and detailed tests have shown that standard jitter
corrections do not improve the data calibration.  Calibration of
fringe data was performed by interspersing target observations with
those of unresolved calibrators; basic information on the calibrator
stars used is given in Table~\ref{cals_tab}.  We note that although
the calibrator stars are significantly brighter than the target FU~Ori
objects at visual wavelengths where the AO correction is performed,
calibration tests \citep[described in][]{rla2005b} have shown that no
systematic effects are introduced.  Further details on the data
reduction and calibration procedures may be found in
\citet{colavita2003} and \citet{swain2003}\footnote{As well as in a
series of technical memos at the Michelson Science Center ({\tt
http://msc.caltech.edu/KISupport/})}.

The calibrated V$^2$ results appear in Table~\ref{visresults_tab}
along with the projected baseline ($u,v$), date of each
observation and calibrator information.  The V$^2$ errors reported in
Table~\ref{visresults_tab} only include statistical errors.  Internal
data quality checks have established a conservative upper limit to the
systematic error $\Delta\,V^2=0.05$, which we add in quadrature and
include in all model fitting (and figures) presented in this paper.

\section{Results}

\subsection{Characteristic Sizes of the Near-Infrared Emission} \label{sizes_sec}

As can been seen in Table~\ref{visresults_tab}, all three sources are
clearly resolved by the KI (calibrated $V^2 < 1.0$).  Since the
complex visibility is related to the source brightness by a Fourier
transform relation \citep[see e.g.][]{goodman1985}, we can use a
simple a priori representation of the object morphology and fit to the
measured visibility amplitudes in order to extract source parameters
of astrophysical value. This approach has the merit of providing
fundamental new information about the NIR emission (size scale) that
is independent of (often poorly constrained) details of specific
physical models.

We use two plausible geometric representations for the NIR brightness:
a Gaussian and a uniform ring. The parameters to be fit are the
Gaussian full width at half-maximum (FWHM) and the ring diameter
respectively (the ring thickness is poorly constrained by measurements
in the first lobe of the visibility function, and following previous
work -- e.g. \citet{jdm2005} -- we adopt a 20\% fractional ring
width).  These results are given in Table~\ref{sizes_tab}. We note
that in our sample, the only object previously resolved at these
scales is V1057~Cyg, for which \citet{wilkin2003} similarly derive a
Gaussian diameter that is larger than ours but within errors.

In the next sections we test the detailed predictions of accretion
disk models, against both our new visibility measurements and
reconstructed broadband SEDs.

\subsection{Constructing the SEDs} \label{sed_sec}

It is not the purpose of this paper to perform detailed SED modelling,
nor do our main conclusions depend on it. However, since FU~Orionis
objects are by definition photometrically variable, we have taken some
care to construct meaningful (i.e. as near-contemporaneous as
possible) SEDs from existing data, as follows:

{\bf V1057~Cyg:} We use near-contemporaneous UBVR data provided by one
of us (Melnikov) and obtained in July and August~2002 at the Maidanak
High Altitude Observatory as part of long term monitoring campaign of
young stars (the ROTOR project, \citet{herbst1999,clarke2005}). I-band
data is interpolated between \citet{kopatskaya2002} and photometry
obtained by one of us (Tannirkulam) at the MDM observatory in
November-December~2004.  The NIR JHK photometry is from
2MASS. \citet{abraham2004} show a $\times 2$ fading at NIR wavelengths
between 1983 and 1996; assuming a constant rate of fading, this
represents only a 0.08~mag fading between the 2MASS (June~2000) and KI
epochs (June, October~2002), un-important for our purposes. We
nevertheless augment the 2MASS errors by 0.1~mag to allow for this
level of variation, as well as flickering known to exist in similar
systems \citep[see e.g.][]{ancker2004,kenyon2000}.  Thermal-IR data is
from ISOPHOT as presented by \citet{abraham2004}.
  
{\bf V1515~Cyg:} Same as above, except that: (1) no I$-$band data is
available, and (2) given the large ISOPHOT beam and beam confusion
issues discussed for this object by \citet{abraham2004}, we use
long-wavelength fluxes from the aperture photometry in
\citet{kenyon1991a}.  We note that the level of NIR fading reported by
\citet{abraham2004} is much less significant even than for v1057~Cyg.
The epoch of Maidanak UBVR data is June~2003; the epoch of JHK 2MASS
data is November~1998.

{\bf Z~CMa-SE:} Composing an SED for the FU~Ori component in this
system is further complicated by the fact that most existing
photometry is for the total system. For UBV, we use the
near-contemporaneous Maidanak photometry (epoch January~2004), given
that at these wavelengths the FU~Ori component is thought to dominate
the total flux \citep[see e.g.][]{thiebaut1995}. An estimate of the
R-band flux is obtained from the Maidanak photometry using the flux
ratio of \citet{thiebaut1995}. Infrared J \& H fluxes are estimated by
combining the 2MASS total fluxes (epoch December~1998) and the flux
ratios from the spatially resolved observations of \citet{rmg02}
(epoch January~2001). At longer wavelengths, the infrared fluxes are
from the discovery observations of \citet{koresko1991} (epoch
Novemeber~1990). The K-band flux has been dimmed by 0.2~mag in order
to account for the fading observed by \citet{ancker2004}.

\subsection{Accretion Disk Models} \label{models1_sec}

Accretion disk models have proved successful at reproducing many of
the observational properties of FU~Orionis objects \citep[][and
references therein]{hartmann1985,hartmann1996}.  A key prediction of
accretion disk models is the shape of the broadband SEDs, and
agreement with observations provide one of the strongest lines of
support for the disk hypothesis.  For a steady Keplerian disk that is
optically thick and geometrically thin, its radial temperature profile
can be parametrized as \citep{shakura1973,lynden1974}: $T(r) = T_{max}
\cdot (r/R_{\star})^{q=-3/4} \cdot [1-(R_{\star}/r)^{1/2}]^{1/4}$, for
$r > 1.36 \, R_{star}$, and $T(r) = T_{max}$ between this radius and
the stellar surface. The maximum temperature in the disk ($T_{max}$)
is related to the stellar mass and disk accretion rate, and good SED
fits are obtained with relatively high values of this parameter
$\sim 5000-7000$~K, together with stellar radii normal for low-mass young
stars $2-4 R_{\sun}$. Under these conditions, the disk flux far
outshines the central star\footnote{We note that SED solutions with
much lower $T_{max}$ and larger $R_{\star}$ are possible, but require
essentially all the U $\rightarrow$ R flux to arise in the stellar
photosphere, which is in conflict with the measured line profiles.}
at all wavelengths and the model has relatively few parameters. If the
disk inclination can be reasonably well constrained by other means,
then the extinction $A_V$, stellar radius $R_{\star}$ and maximum disk
temperature $T_{max}$ are all well constrained by the shape and level
of the SED.  Once a model solution is thus obtained, the corresponding
visibilities can be computed and a straightforward comparison to the
KI data be made.

Another characteristic of these single power-law models is that,
except for inclinations effects (believed to be low for our sources,
e.g. \citet{kenyon1988}), flux and ``size'' at a given wavelength are
simply related. This scaling, and comparison with the FU~Ori {\em
Palomar Testbed Interferometer} (PTI) observations of
\citet{malbet1998}, immediately reveals that our targets are
over-resolved with respect to expectations from the disk model.
Indeed, our observations of V1057~Cyg and V1515~Cyg are approximately
consistent with the visibility curve for FU~Ori in \citet{malbet1998},
corresponding to a disk solution which also fits its SED. However,
V1057~Cyg and V1515~Cyg are $\sim \times 38$ and $\times 8$ {\em
fainter} than FU~Ori, respectively. Therefore, a disk model that fits
the SEDs of V1057~Cyg and V1515~Cyg will clearly under-estimate their
visibilities.

Following previous workers, our disk is represented by a series of
concentric annuli. At each radius, the annulus radiates as a Kurucz
\citep{kurucz1979} supergiant photosphere of effective temperature
equal to the disk temperature at that radius, unless the temperature
is lower than 3500~K, in which case we use blackbody emission. We use
solar metallicity and a value of the surface gravity appropriate for
the temperature at each radius.  Model SEDs are reddened using the law
of \citet{mathis1990}, using a total-to-selective extinction ratio
$R_V=3.1$.  The KI data were obtained essentially at a single baseline
position angle and therefore can not constrain the disk inclination.
Moreover, for V1057~Cyg and V1515~Cyg, the rotational velocities
measured from line broadening are relatively low, such that unless the
stars have unreasonably low mass, the inclinations must be low
\citep[$i<30 \degr$,][]{kenyon1988}.  Therefore we consider mainly
face-on disks in our fits, but explore the effects of disk
inclination.

\begin{figure}
\begin{center}
\includegraphics[angle=90,scale=0.7]{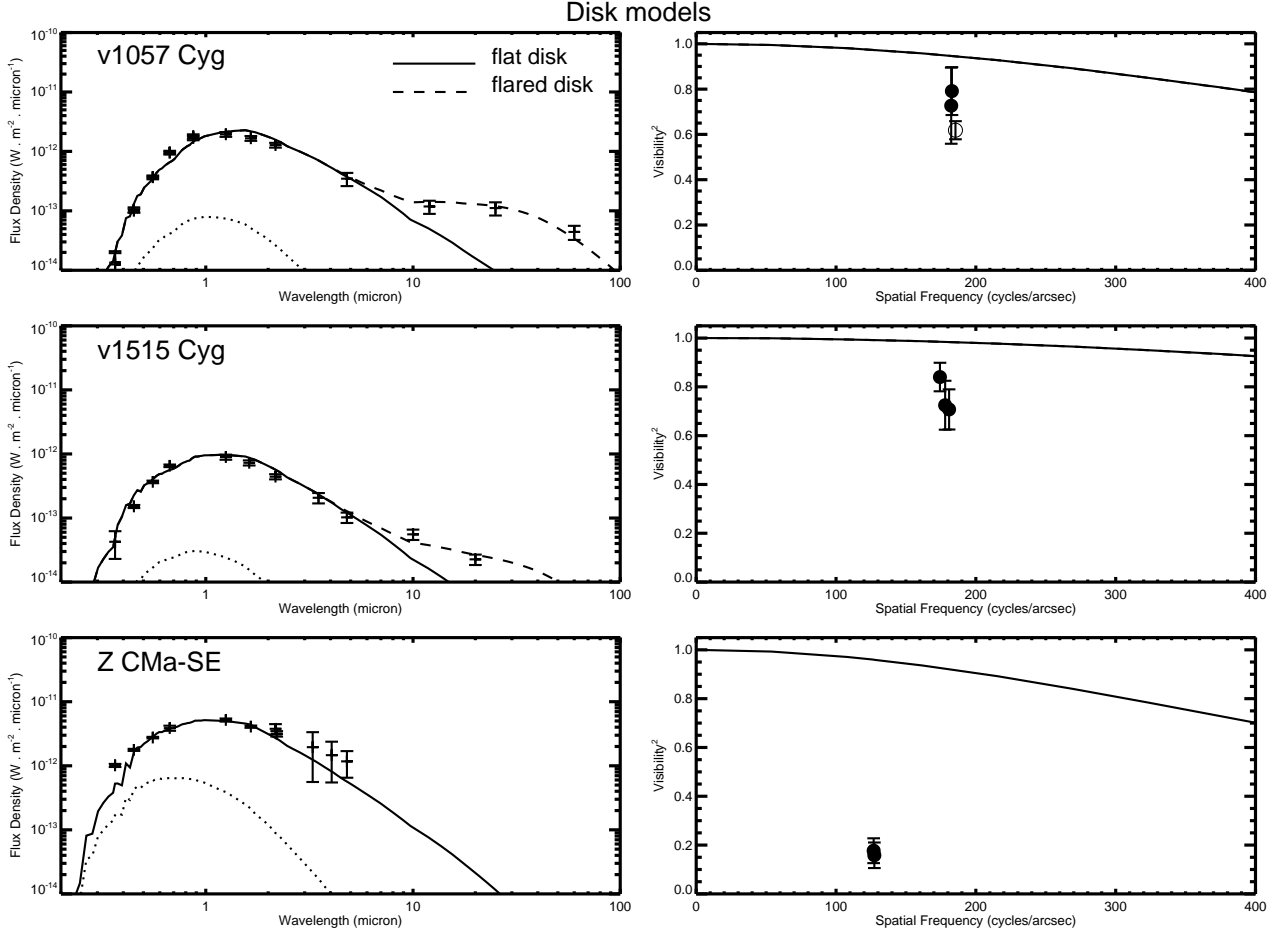}
\caption{
Data and disk models results, SED and visibility amplitudes in left
and right panels respectively.  For V1057~Cyg we include the visibility data
obtained at the PTI by \citet{wilkin2003} (open symbol).  The models
correspond to best-SED-fits and are calculated for the parameters of
Table~\ref{params_tab}.  The solid line shows the model for a flat
disk, the dashed line shows the model for a flared disk.  For the flat
disk model, the fitting procedure uses only SED data with $\lambda <
10 \mu m$; while fitting the flared disk model uses the whole SED.  In
the visibility panels the two lines overlap because the flared disk
regions do not contribute significant K-band flux. The dotted line in
the SED panels is the stellar photosphere. For Z~CMA-SE, we consider
only the flat disk model, given the lack of resolved long wavelength
photometry.
\label{fig1}}
\end{center}
\end{figure}

The solutions found from fitting the SEDs are similar to those found
by previous workers, and are shown in Table~\ref{params_tab}. The data
and disk models are shown in Figure~\ref{fig1}. For V1057~Cyg, in this
and subsequent visibility figures, we include (open symbol) the data
obtained at the PTI by \citet{wilkin2003}.  However, due to the very
different FOVs of the KI and PTI instruments, and given the possible
effect of large scale emission in these objects (see
$\S$~\ref{models2_sec}) we choose not to include this datum in our
fitting exercises.  In the visibility panels it can be clearly seen
that geometrically flat accretion disks (i.e. single power law with
$q=-0.75$) significantly overestimate the visibility amplitudes at
K-band, because the relatively steep temperature profile results in
the NIR emission coming from regions of the disk that have too small
radii.

As is well known, and as is also apparent from the SED panels of
Figure~\ref{fig1}, for some FU~Orionis objects (such as V1057~Cyg and
V1515~Cyg) longward of $\sim 10 \mu m$ the infrared fluxes far exceed
the flat disk predictions. Flared disks having a flatter temperature
profile have been proposed as one solution to explain the
long-wavelength SED \citep[see e.g.][]{kenyon1991b,lachaume2004}. We
have approximated this model by specifying a double-power law for the
temperature profile in the disk. Following \citet{kenyon1991b}, we use
Figure~4 of their paper, and select the case of most extreme flaring
($H_0=0.2$, this parameter determines the disk height profile as a
function of radius, $H_d(r)/R_{\star} = H_0 \cdot
(r/R_{\star})^{9/8}$), to approximate their detailed temperature
profile with two slopes: $q = -0.75$ for $r < 100 \cdot R_{\star}$ and
$q_2 = -0.45$ outside this radius.  Figure~\ref{fig1} shows that for
flared disks the long wavelength SED is now well reproduced; however
the K-band visibilities are indistinguishable from the non-flared
case, simply because the flared disk regions are too cool to thermally
emit significant NIR radiation.

For Z~CMa-SE, there exists no resolved long wavelength photometry, and
therefore we consider only the flat disk model. This source is much
more resolved than the other two, at levels very difficult to explain
with thermal disk emission only.

\begin{figure}
\begin{center}
\includegraphics[angle=90,scale=0.7]{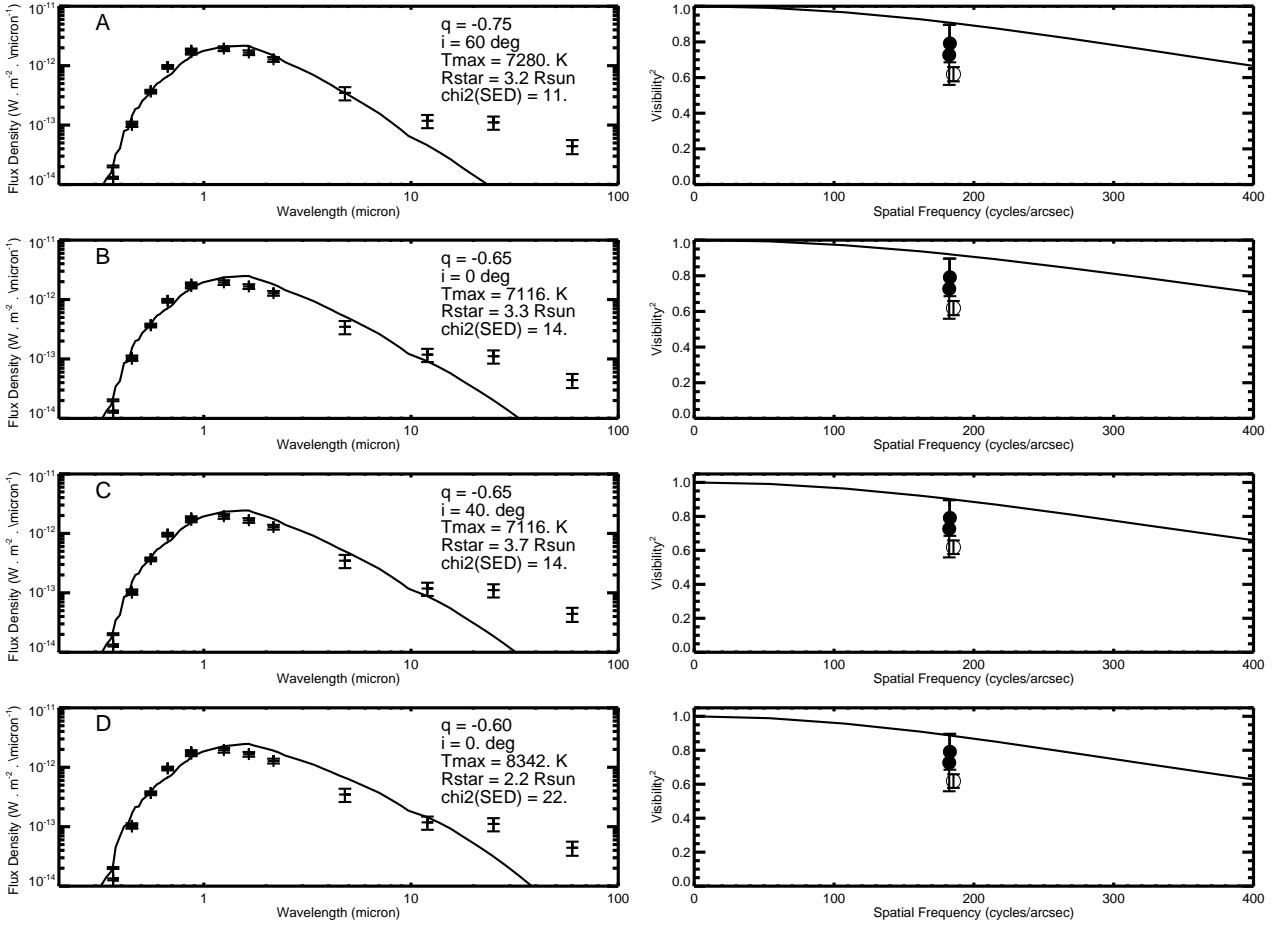}
\caption{
{\bf V1057~Cyg}: Effect of disk inclination (A) and different values
of the temperature radial exponent (B $\rightarrow$ D) for the
inner-disk. These models may be compared with the V1057~Cyg best fit
(zero inclination) of Figure~\ref{fig1}, for which we have the
reference reduced $\chi^2_{SED} = 11$. {\bf Panels A:} show that
although a $60 \degr$ inclination reproduces the SED equally well, it
is the minimum inclination required to just become consistent with the
upper limit of the $V^2$ data.  {\bf Panels B:} show that $q=-0.65$
produces a marginally acceptable SED fit, but is not sufficient to
touch the $V^2$ upper bound.  {\bf Panels C:} are also for $q=-0.65$,
and show that additionally including inclination is able to reproduce
the $V^2$ upper bound, but a relatively large value of $40 \degr$ is
required.  {\bf Panels D:} illustrate that $q=-0.6$ reproduce the
$V^2$ upper bound, even for zero inclination, but results in a
significantly poorer SED fit.  In all cases, SED data longward of $10
\mu m$, although plotted, is not included in the calculation of the
$\chi^2_{SED}$, since here we only attempt to model the inner disk
regions where the K-band flux arises. For V1057~Cyg we include the visibility data
obtained at the PTI by \citet{wilkin2003} (open symbol).
\label{fig1b}}
\end{center}
\end{figure}

We have explored two additional possibilities to attempt to
explain the low visibilities measured by KI using purely thermal disk
emission: disk inclination and a non-standard temperature profile
($q \neq -0.75$) for the inner-disk, as follows:

The SEDs can be well fit for non-zero inclinations, provided the flux
lost to surface area is recovered by increasing $R_{\star}$; which
also has the desired effect of resulting in lower visibilities (a
larger NIR source) if the KI baseline position angle also happens to
be aligned with a long direction of the elliptical brightness.
However, we find that, even allowing for the most favorable disk
orientation, the inclinations needed to match the KI data are very
large, $i \sim 60 - 85 \degr$, well outside the upper limit inferred
from line broadening for typical young star masses range, as mentioned
above. We illustrate this difficulty for V1057~Cyg in the top panel of
Figure~\ref{fig1b}, and similar conclusions apply to V1515~Cyg.

Relaxing the assumption that $q = -3/4$ and allowing a flatter profile
in the inner disk would also result in larger NIR emission regions.
Exploring this departure from the canonical temperature profile
exponent is a reasonable approach, given that (a) the decaying and
flickering light curves show that the disk is not exactly
steady-state, and (b) the absence of the expected boundary layer
emission \citep{kenyon1989} calls into question the radial temperature
profile near the inner disk edge.  We find however that matching the
visibilities would require $q \ge -0.6$, and for such exponents the
SED can not be satisfactorily fit, because of the sensitivity of its
infrared slope to the temperature profile exponent. Allowing a free
q-parameter {\it and} the disk to be inclined does not provide a
satisfactory solution either, simply because $q \sim -0.65$ is the
flattest exponent allowed by the NIR SED slope, and for such exponents
large disk inclinations would again be needed to match the
visibilities. We illustrate these difficulties for V1057~Cyg in the
bottom three panels of Figure~\ref{fig1b}, and similar conclusions
apply to V1515~Cyg.

Finally, we note that allowing the disk inner radius to be a free
parameter ($r_{inner} >> R_{\star}$ can formally result in acceptable
simultaneous fits to the SEDs and visibilities, but we believe are not
physically viable for our targets (contrary to the case of
Herbig~Ae/Be and T~Tauri systems).  Allowing an inner disk radius
several times larger than $R_{\star}$ (for example $10 \times
R_{\star}$ for V1057~Cyg), and keeping the same $T_{\max}$ to preserve
the SED shape, results in NIR emission originating at larger radii
(comparable to the ring radii of $\S$~\ref{sizes_sec}), and lower
visibilities in agreement with the observations. However, to preserve
the SED flux levels, the disk needs to be inclined by very large
amounts ($86 \degr$ for V1057~Cyg), in contradiction to the limits
derived by \citet{kenyon1988}. Moreover, physical disk models
\citep{clarke1990,hartmann1993,bell1994} indicate that for the high
accretion rates of FU~Ori systems, it is very implausible that a
central optically thin disk region could exist; and the emission lines
that would be expected if it did are not observed.

\subsection{Additional Model Components} \label{models2_sec}

The difficulties encountered in the previous section in using standard
disk models to simultaneously reproduce the observed SEDs and
visibilities may indicate that these models are inadequate for
describing the NIR emission.  Alternatively, additional source
components could exist which are responsible for the low measured
visibilities.

\subsubsection{Extended Structure} \label{models2a_sec}

\begin{figure}
\begin{center}
\includegraphics[angle=90,scale=0.7]{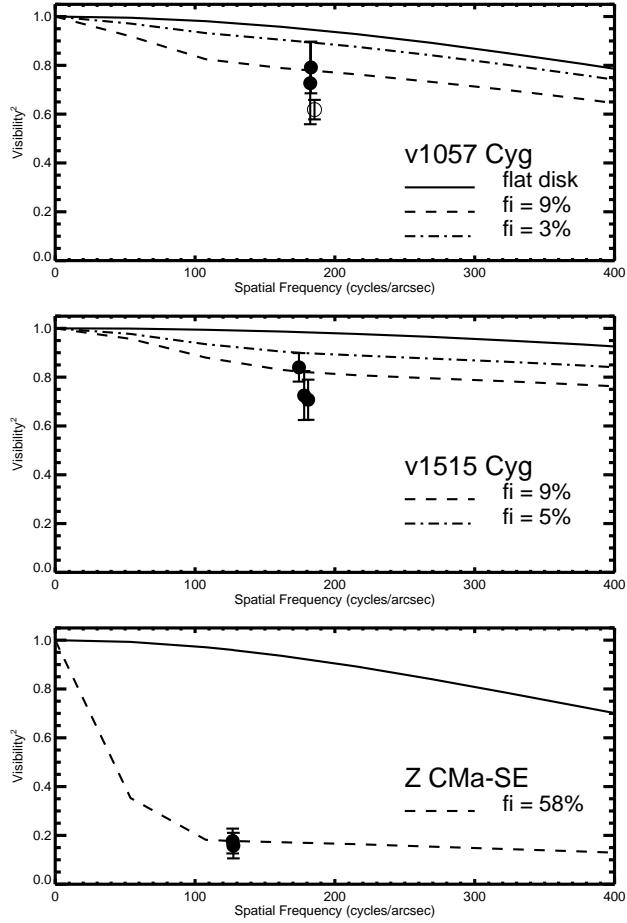}
\caption{ Adding incoherent fluxes to the best-SED-fit disk models of
Figure~\ref{fig1} (solid line).  The dashed lines show how the
visibilities are reduced to the desired levels by adding the indicated
amounts of incoherent (completely resolved) flux to the disk models.
The size of the region where this incoherent flux arises is not
constrained by the observations, but for display purposes it has been
included in the model as a Gaussian component of FWHM equal to minimum
value which is completely resolved at the data spatial frequencies:
4~AU, 4~AU and 10~AU for V1057~Cyg, V1515~Cyg and ZCMa-SE
respectively.  The dotted line in the SED panels is the stellar
photosphere. For V1057~Cyg we include the visibility data obtained at
the PTI by \citet{wilkin2003} (open symbol).
\label{fig2}}
\end{center}
\end{figure}

A possible explanation is that the measured visibilities are not
entirely due to compact NIR thermal emission in a flat or flared disk,
but that small but non-negligible fraction of the total flux arises in
larger scale structure within the 50~mas FOV of the interferometer
(25~AU and 50~AU at $d=500,1000$~pc respectively). The size of this
region and the flux it contributes are degenerate when modelling its
contribution to the visibility.  However, a {\it minimum} flux can be
inferred if we assume that it is completely incoherent
(i.e. completely resolved by the interferometer), which for the
spatial frequencies sampled and the distances to the three objects
corresponds to size scales $\gtrsim$~4~AU, 5~AU, and 10~AU, for
V1057~Cyg, V1515~Cyg and ZCMa-SE respectively.  The needed amount of
incoherent flux can be analitically estimated as follows. For an
object consisting of a central compact object (the NIR disk, we
neglect the stellar flux) plus an extended component, the measured
visibility (at a given spatial frequency) is: $V_{measured} = (F_C
\cdot V_C + F_E \cdot V_E)/F_T$; where the subscripts in the flux
($F$) and visibility ($V$) of each component denote:
$C=\mbox{Compact}$, $E=\mbox{Extended}$ and $T=\mbox{Total}$.  In the
case that the extended component is completely resolved, we have $V_E
= 0.0$, and the minimum incoherent flux fraction is given by: $f_i =
F_E/F_T = 1.0 - V_{measured}/V_C$. Table~\ref{params_tab} summarizes
the values for $f_i$ obtained for the cases that the compact object is
unresolved ($V_C = 1.0$) and, more meaningfully, the case that the
compact object is the accretion disk considered in the previous
section, and is resolved at the data spatial frequencies by the
amounts predicted by the best-SED-fit models (Table~\ref{params_tab}
and Figure~\ref{fig1}).  The effect of this additional component on
the visibility curves is shown in Figure~\ref{fig2}, where the ranges
of $f_i$ shown correspond to the values needed to reproduce the
weighted mean visibility measured, and visibility data upper limits
given the $+1 \, \sigma$ errors.

\subsubsection{Physical Origin of the Extended Emission}

Given the low thermal temperatures expected for any circumstellar
material located at the scales considered above, physically this
additional emission would correspond to scattering of thermal disk
K-band photons by circumstellar material (at least for V1057~Cyg and
V1515~Cyg, the fraction of incoherent flux derived above for ZCMa-SE
is prohibitely high for a scattering origin).  Our observations do not
directly constrain the precise nature of the putative incoherent flux
source, reasonable hypothesis are that scattering occurs in the upper
atmosphere of the outer disk, or in envelope material.

\citet{kenyon1991b} first showed that thermal emission in infalling
envelopes surrounding the V1057~Cyg and V1515~Cyg disks provide good
fits to the long wavelength excess fluxes in these systems.  The
putative envelopes have evacuated cones along the poles, cleared by
bipolar outflows, which result in the relative un-obscured views
towards the central objects, as observed. Moreover, these authors
favor the envelope hypothesis over flaring disks, due to the large
degree of flaring that would be required to reproduce the
long-wavelength SEDs, and the wavelength dependent photometric decay
curves. For T~Tauri objects with flat SEDs, \citet{natta1993} have
also proposed the existence of envelopes (tenous in that case) which
heat the outer disk regions by scattering stellar photons onto it.

The envelopes considered by \citet{kenyon1991b} for V1057~Cyg and
V1515~Cyg are characterized by $\sim 7$~AU inner radii and
temperatures at these radii of $\sim$~few~100~K; and it is therefore
easy to see that adding such an envelope to the disk has no effect on
the K-band visibilities if only thermal emission is considered; as
pointed out by these authors the emission below 10~$\mu m$, and
certainly at 2~$\mu m$, is completely dominated by the disk. Moreover,
the $\sim 7$~AU inner radii imply that this component would indeed be
completely resolved by the KI. Therefore, the basic considerations
presented in $\S$~\ref{models2a_sec} provide the amounts of flux that
the envelope would need to scatter at K-band to reproduce the KI
observations.

These are relatively large amounts of scattered K-band flux,
originating in a relatively small FOV.  However, building upon the
initial models of \citet{kenyon1991b}, more refined models of
V1057~Cyg and V1515~Cyg are being explored and compared with new data
from the Infrared Spectrograph on board the Spitzer Space Telescope,
which involve infall to the disk at about 10~AU \citep{green2005}.
Independent of model details, the long-wavelength excesses above what
is expected for the disk are approximately 20\% of the apparent total
disk luminosity for V1057~Cyg and 12\% of the total apparent
luminosity of V1515~Cyg.  If the effective scattering albedo of the
envelope in the K band is of order of 0.5, then the envelopes invoked
to explain the infrared excesses longward of $\sim 10 \mu$m could be
responsible for a scattered flux of 6$-$10\% within the FOV of the
interferometer.  This is in reasonable agreement with the
observational requirements, given the uncertainties and likely complex
geometry, implying that the scattering envelopes of V1057~Cyg and
V1515~Cyg can in fact account for the K-band visibilities.

\subsubsection{Stellar Companion}

Dynamical interactions with a stellar companion have been invoked to
explain the origin of the accretion rate outbursts in FU~Ori objects
\citep{bonnell1992}, and in fact several objects of the class are
known to have ``wide'' (tenths of arcsec separations) companions.
Motivated by the recent discovery of such a wide companion (separation
$\sim 0.5 \arcsec$) to FU~Ori itself \citep{wang2004},
\citet{reipurth2004} propose that the FU~Ori phenomenon corresponds to
the formation of binary systems by break up of small nonhierarchical
systems. For FU~Ori itself, this scenario predicts a third close
companion, at separation $\sim 10$~AU, perhaps corresponding to the
interferometric signature detected by \citet{malbet2005}.

A binary system produces a sinusoidal visibility amplitude curve as a
function of projected baseline, with period given by the (inverse of)
angular separation and an amplitude given by the flux ratio between
the two components. Given our limited data set (essentially a single
spatial frequency was measured for each object), and the sinusoidal
response, many values of the binary parameters (flux ratio and
companion location) result in visibility functions that reproduce our
measurements equally well, and therefore a fitting excercise is not
very informative. However, for reference, interesting bounds on the
family of possible solutions are as follows. The measured visibility
values imply a {\em maximum} flux ratio between the binary components
of about: 20:1, 20:1 and 2.5:1 for V1057~Cyg, V1515~Cyg and Z~Cma-SE
respectively. For this maximum value of the flux ratio, the {\em
minimum} angular separations are about: 1.0, 1.0 and 1.7~mas for
V1057~Cyg, V1515~Cyg and Z~Cma-SE respectively. For equal flux
components, the minimum separations become: 0.4, 0.4 and 1.3~mas for
V1057~Cyg, V1515~Cyg and Z~Cma-SE respectively. We note that for these
estimates we have considered a binary model in which the ``primary''
component is again the partially resolved best-SED-fit disk of
$\S$~\ref{models1_sec}.

\section{Discussion and Conclusions} \label{conclusions}

We have presented new observations which spatially resolve the NIR
emission of three FU~Orionis objects on mas scales, a first for two of
the objects (V1515~Cyg and Z~CMa-SE). The spatial scales for NIR
emission are surprisingly large, compared to predictions of simple
accretion disk models.  This indicates that the accretion models
require modification, or that additional model components are
required.

As is well known, the infrared SED slopes sensitively probe the radial
temperature profile in the disk, and we have shown that while
exponents $q=-0.6$ or flatter would be required to match the measured
visibilities, these solutions are not able to maintain good agreement
with the SED data.  We note however that in this approach, tests of
the validity of small variations to the disk temperature profile are
limited by the accuracy of the SED photometry adopted, and therefore
follow-up studies should ideally be performed using contemporaneous
photometry of uniform quality.

It is interesting to contrast our results with the case of the
prototype for the class, FU~Orionis, which has only modest long
wavelength fluxes in excess of thermal emission in a flat accretion
disk, and for which the NIR interferometer data does agree well with
this model according to \citet{malbet1998,malbet2005}. We note that
this conclusion rests on the ability to incline the FU~Ori disk by a
relatively large amount ($i \sim 50-60 \degr$), which in this case does
not imply unreasonably low stellar masses given the measured
rotational velocities \citep{kenyon1988}.

Given the existing indications for dense envelopes around V1057~Cyg
and V1515~Cyg, we hypothesize as a natural explanation for the low
measured visibilities that scattering by material located on 10s of AU
scales adds incoherent flux in the interferometer FOV, diluting the
visibilities from the relatively compact inner disk. We note that this
mechanism may also explain the somewhat larger size (Gaussian
FWHM~=~$1.36 \pm 0.07$~mas) found for V1057~Cyg by \citet{wilkin2003},
since the Palomar Testbed Interferometer has a $1 \arcsec$ FOV,
$\times 20^2$ larger than that of the KI. A full calculation of FOV
effects in specific scattering models is beyond the scope of this
paper, we simply note that considering the PTI measurement alone, the
minimum incoherent flux fraction becomes $f_i = 21 \pm 5$~\%.

An important consequence of this interpretation is that it compromises
the notion to use FU~Orionis objects as {\em straightforward} test
cases of accretion disk theories, at least for most of the better
known candidate targets. Indeed, if the disk is not isolated,
multi-baseline and multi-wavelength interferometer observations must
be used to better discriminate between competing models.  In
particular, the hypothesis that scattering is at the origin of the low
visibilities measured is testable using measurements in different NIR
bands: at shorter wavelengths were scattering is more efficient
(e.g. J or H bands) the characteristic sizes should be larger; while
thermal disks predict the opposite trend of larger sizes at longer
wavelengths (and with a detailed dependence that probes the flatness
of the temperature profile). These multi-wavelength observations are
already possible at existing interferometers (e.g. KI and the AMBER
instrument on the VLTI, \citet{petrov2000}) or upcoming instruments
(e.g. the MIRC combiner at CHARA, \citet{monnier2004}).  For objects
with evidence for dense envelopes, self-consistent radiative transfer
codes that include scattering should also be used in order to include
this effect in both the SEDs and visibilities for realistic geometries
and dust properties.

The exciting possibility that all three objects are instead resolved
due to the presence of stellar companions, sometimes invoked to
explain the outburst themselves, is also testable with follow-up
interferometer observations with improved spatial frequency coverage.
Given the KI visibility calibration accuracy (which translates into a
100:1 maximum detectable binary contrast), the telltale and
un-ambiguous sinusoidal signatures would be easily detected even for
the most pessimistic values of the binary contrast derived above.

\acknowledgments

The authors wish to thank the all members of the Keck Interferometer
development team (JPL, MSC, WMKO) whose dedicated efforts made this
``shared-risk'' commissioning science possible.  We also thank
Dr. Francis Wilkin for providing the PTI data for V1057~Cyg.  This
material is based upon work supported by NASA under JPL Contracts
1236050 \& 1248252 issued through the Office of Space Science.  Data
presented herein were obtained at the W.M. Keck Observatory from
telescope time allocated to the National Aeronautics and Space
Administration through the agency's scientific partnership with the
California Institute of Technology and the University of
California. The Observatory was made possible by the generous
financial support of the W.M. Keck Foundation.  This research has made
use of the SIMBAD database, operated at CDS, Strasbourg, France. This
publication makes use of data products from the Two Micron All Sky
Survey (2MASS), which is a joint project of the University of
Massachusetts and the Infrared Processing and Analysis
Center/California Institute of Technology, funded by the National
Aeronautics and Space Administration and the National Science
Foundation.  This work has made use of services produced by the
Michelson Science Center at the California Institute of Technology.
The authors wish to recognize and acknowledge the very significant
cultural role and reverence that the summit of Mauna Kea has always
had within the indigenous Hawaiian community.  We are most fortunate
to have the opportunity to conduct observations from this mountain.

\clearpage



\begin{deluxetable}{lccccccccc}
\rotate
\tablecolumns{10}
\tablewidth{0pc}
\tablecaption{Basic Properties of Targets. \label{targets_tab}}
\tablehead{
\colhead{Name} & \colhead{RA} & \colhead{Dec} & \colhead{V} & \colhead{J} & \colhead{H} & \colhead{K} &%
\colhead{d} & \colhead{Adopted} & \colhead{SED} \\
 & (J2000) & (J2000) &%
(mag)\tablenotemark{a} & (mag)\tablenotemark{a} & (mag)\tablenotemark{a} & (mag)\tablenotemark{a}%
& (pc) & $T_{\star}$(K)\tablenotemark{b} &  photometry references\tablenotemark{c}
}
\startdata
V1057 Cyg & 20 58 53.7 & $+$44 15 28.4     & 11.7     & 8.0 & 7.0 & 6.2 & 550 \phn $\pm$ 100  (1) & 6200 & (4,5,6,7) \\
V1515 Cyg & 20 23 47.6 & $+$42 12 24 \phn  & 12.4     & 8.9 & 8.0 & 7.4 & 1000 $\pm$ 200 (2) & 5845 & (5,7,8) \\
Z CMa-SE  & 07 03 43.2 & $-$11 33 06.2     & \phn 9.8 & 6.5 & 5.2 & 3.8 & 1150           (3) & 6360 & (5,7,9,10) \\
\enddata
\tablenotetext{a}{As discussed in the text, these objects are by
defition highly variable.  The V and infrared magnitudes in this table
are from SIMBAD and 2MASS respectively, and are intended to be merely
representative. See the text and references in this table for a
description of how the SED used in the data analysis have been
constructed.  For Z~CMa the magnitudes given in this table are for the
total system.}
\tablenotetext{b}{
The true photospheric spectral types are not known as there are no
pre-outburst spectra of these objects, these values are typical of
what is quoted in the literature. The exact photospheric effective temperatures are
however inconsequential to the analysis given the negligible stellar
flux contribution in the models considered.}
\tablenotetext{c}{As indicated in the text, we use
near-contemporaneous UBVR photometry obtained at Maidanak Observatory
(ROTOR program). In this column we provide the relevant references for
additional literature photometry used to construct our SEDs.}
\tablerefs{
(1) \citet{straizys89};
(2) \citet{racine68};
(3) \citet{herbst78};
(4) \citet{kopatskaya2002};
(5) 2MASS All-Sky Point Source Catalog;
(6) \citet{abraham2004};
(7) \citet{ancker2004};
(8) \citet{kenyon1991a};
(9) \citet{rmg02};
(10) \citet{koresko1991};
}
\end{deluxetable}


\clearpage


\begin{deluxetable}{lccccc}
\tablecolumns{6}
\tablewidth{0pc}
\tablecaption{Calibrator Stars Information. \label{cals_tab}}
\tablehead{
\colhead{Name} & \colhead{V} & \colhead{J} & \colhead{K} & \colhead{Spectral Type} & \colhead{UD diameter (mas)}
}
\startdata
HD~199178  & 7.2 & 5.7 & 5.2 & G2V & 0.44 $\pm$ 0.07 \\
HD~192985  & 5.8 & 5.0 & 4.8 & F5V & 0.43 $\pm$ 0.06 \\
HIP~102667 & 8.8 & 6.4 & 5.5 & K2V & 0.49 $\pm$ 0.16 \\
HD~227049  & 9.1 & 7.0 & 6.3 & K2III & 0.30 $\pm$ 0.10 \\
HD~332518  & 9.2 & 6.9 & 6.2 & K5V   & 0.38 $\pm$ 0.08 \\
HD~52919   & 8.3 & 6.4 & 5.7 & K5V   & 0.42 $\pm$ 0.09 \\
HD~48286   & 7.1 & 5.9 & 5.5 & F7V   & 0.35 $\pm$ 0.08 \\
HD~60491   & 8.2 & 6.5 & 6.0 & K2V   & 0.33 $\pm$ 0.05 \\
\enddata

\end{deluxetable}


\clearpage


\begin{deluxetable}{lcccccc}
\rotate
\tablecolumns{7}
\tablewidth{0pc}
\tablecaption{Calibrated Visibility Amplitude Data. \label{visresults_tab}}
\tablehead{
\colhead{Name} & \colhead{U.T. Date} & \colhead{Julian Date} & \colhead{u (m)} & \colhead{v (m)} & \colhead{$V^2$} & \colhead{Calibrators}
}
\startdata
V1057 Cyg & Jun 27 2002 & 2452453 & 44.519 & 67.770 & 0.727 $\pm$ 0.161 & HD199178, HD192985, HIP102667 \\
V1057 Cyg & Oct 24 2002 & 2452572 & 43.427 & 70.711 & 0.791 $\pm$ 0.093 & HD199178, HD199998 \\ 
V1515 Cyg & May 21 2003 & 2452781 & 54.973 & 55.878 & 0.840 $\pm$ 0.030 & HD227049, HD332518 \\
V1515 Cyg & May 21 2003 & 2452781 & 53.245 & 59.776 & 0.725 $\pm$ 0.087 & HD227049, HD332518 \\
V1515 Cyg & May 21 2003 & 2452781 & 51.096 & 63.262 & 0.707 $\pm$ 0.065 & HD227049, HD332518 \\
Z CMa-SE  & Apr 03 2004 & 2453099 & 24.195 & 51.929 & 0.158 $\pm$ 0.016 & HD48286, HD52919, HD60491 \\
Z CMa-SE  & Apr 03 2004 & 2453099 & 23.753 & 51.887 & 0.177 $\pm$ 0.009 & HD48286, HD52919, HD60491 \\
\enddata
\tablecomments{This table shows only the statistical $V^2$ error, 
as discussed in $\S$~\ref{obs_sec} an additional systematic error of 0.05 is added (in quadrature) 
to all data for fitting purposes and the total error also is shown in all figures.}
\end{deluxetable}


\clearpage


\begin{deluxetable}{lccccc}
\tablecolumns{6}
\tablewidth{0pc}
\tablecaption{Characteristic Near-Infrared Sizes. \label{sizes_tab}}
\tablehead{
\colhead{} & \multicolumn{2}{c}{Gaussian FWHM} & \colhead{} & \multicolumn{2}{c}{Ring Inner Diameter} \\
\cline{2-3}  \cline{5-6} \\
\colhead{Name} & \colhead{(mas)} & \colhead{(AU)} & \colhead{} & \colhead{(mas)} & \colhead{(AU)} 
}
\startdata
V1057 Cyg & 1.04 $\pm$ 0.23 & 0.57 $\pm$ 0.16 & & 1.11 $\pm$ 0.24 & 0.55 $\pm$ 0.15 \\
V1515 Cyg & 1.05 $\pm$ 0.12 & 1.05 $\pm$ 0.24 & & 1.13 $\pm$ 0.12 & 1.13 $\pm$ 0.20 \\
Z CMa-SE  & 3.94 $\pm$ 0.24 & 4.53 $\pm$ 0.48 & & 3.81 $\pm$ 0.18 & 4.38 $\pm$ 0.21 \\
\enddata

\end{deluxetable}


\clearpage


\begin{deluxetable}{lcccccccccccc}
\rotate
\tablecolumns{13}
\tablewidth{0pc}
\tablecaption{Model parameters. \label{params_tab}}
\tablehead{
\colhead{}    &  \multicolumn{4}{c}{Flat Disk} &   \colhead{}   & \multicolumn{2}{c}{Flared Disk} 
& \colhead{}   & \multicolumn{2}{c}{Minimum Incoherent Fluxes} \\
\cline{2-5} \cline{7-8} \cline{10-11} \\
\colhead{Name} & \colhead{$q$}   & \colhead{$T_{max}$(K)}    & \colhead{$R_{\star}/R_{\sun}$} & \colhead{$A_V$} &
\colhead{}    & \colhead{$q_2$}   & \colhead{$r_0/R_{\star}$} &
\colhead{}    & \colhead{$f_i(\%) \; (V_C = 1.0)$}   & \colhead{$f_i(\%) \; (V_C = V_{Disk})$} 
}
\startdata
v1057~Cyg & $-$0.75 & 7280. $\pm$ 191. & 2.4 $\pm$ 0.1 & 4.2 & & $-$0.45 & 100.    & &  12 $\pm$ 5 & 9 $\pm$ 5 \\
v1515~Cyg & $-$0.75 & 8014. $\pm$ 280. & 2.1 $\pm$ 0.2 & 3.0 & & $-$0.45 & 100.    & &  10 $\pm$ 2 & 9  $\pm$ 2 \\
Z CMa-SE  & $-$0.75 & 5785. $\pm$ 351. & 7.7 $\pm$ 0.6 & 1.9 & & \nodata & \nodata & &  59 $\pm$ 3 & 58 $\pm$ 3 \\
\enddata

\end{deluxetable}



\clearpage

\clearpage

\clearpage

\end{document}